\documentclass[pra,reprint,showpacs,showkeys]{revtex4-1}
\usepackage{epsf,epsfig,amssymb,graphicx,times}
\usepackage{dsfont,color}
\begin{document}

\title{Sufficiency of quantum non-Gaussianity for discrete-variable quantum key distribution over noisy channels}

\author{Miko{\l}aj Lasota}
\email{Corresponding author. E-mail: lasota@optics.upol.cz}
\author{Radim Filip}
\email{E-mail: filip@optics.upol.cz}
\author{Vladyslav C. Usenko}
\email{E-mail: usenko@optics.upol.cz}
\affiliation{Department of Optics, Palack\'{y} University, 17.\,listopadu 1192/12, 77146 Olomouc, Czech Republic}

\begin{abstract}

Quantum key distribution can be enhanced and extended if nonclassical single-photon states of light are used. We study a connection between the security of quantum key distribution and quantum non-Gaussianity of light arriving at the receiver's detection system after the propagation through a noisy quantum channel, being under full control of an eavesdropper performing general collective attacks. We show that while quantum nonclassicality exhibited by the light arriving at the receiver's station is a necessary indication of the security of the discrete-variable protocols, quantum non-Gaussianity can be a sufficient indication of their security. Therefore, checking for non-Gaussianity of this light by performing standard autocorrelation function measurement can be used for prior verification of the usability of prepare-and-measure schemes. It can play similar role to the prior verification of the quantum correlations sufficient to violate Bell inequalities  for entanglement-based protocols.

\end{abstract}

\maketitle

\section{Introduction}

During more than thirty years since the invention of the famous BB84 protocol \cite{Bennett84} many different discrete-variable quantum key distribution (DV QKD) schemes, utilizing single photons to generate secure keys between their legitimate participants (traditionally called Alice and Bob), have been both proposed theoretically \cite{Ekert91,Bennett92a,*Huttner95,*Bruss98,*Bechmann00,*Scarani04,*Deng04,Hwang03,Barrett05,Acin06,Braunstein12,*Lo12} and realized experimentally \cite{Bennett92b,*Muller95,*Marand95,*Beveratos02,*Zhao06,*Schmitt07,Jennewein00,*Naik00,*Tittel00,Liu13,*Ferreira13}. Comprehensive reviews of this field can be found in \cite{Gisin02,Scarani09}. Although in theory QKD protocols provide one with unconditionally secure way of exchanging information  \cite{Shor00,Kraus05,*Renner05}, in realistic situations their security is typically strongly limited due to various imperfections of the currently available laboratory equipment which can be used for the implementation of these schemes \cite{Lutkenhaus99,Brassard00,Gottesman04}. The most crucial issues for DV QKD protocols in this context are non-zero probability of emitting multiphoton pulses, losses of photons during their propagation between Alice and Bob and errors generated in Bob's detection system \cite{Brassard00}. Usually the primary source of these errors are dark counts, which are randomly generated from time to time in all types of realistic single-photon detectors \cite{Eisaman11}. However they can be caused  also by a few other reasons, \emph{e.g.}\,by the disturbance of photons sent from Alice's source during their propagation to Bob \cite{Gisin92,*Muller97,*Galtarossa05} or by coupling the real signals to noise photons generated by the external sources of light, independent from the legitimate participants of a given QKD protocol \cite{Miao05,*Bonato09}. In particular, channel noise can arise in the dense wavelength-division multiplexing channels due to the crosstalk with strong classical signals or Raman scattering \cite{Eraerds10,*Qi10}.

In general, the aforementioned problems with the setup in real-life scenario may allow a potential eavesdropper (Eve) to hide errors made by her own efforts to gain knowledge on the key created by Alice and Bob, masking them as naturally occurring errors, or to perform more dangerous types of attacks, \emph{e.g.}\,photon-number-splitting (PNS) attacks \cite{Brassard00}, which can be error-free. Thus, in realistic situation in order to preserve at least limited ability to produce a secure key by a given QKD scheme Alice and Bob have to be able to somehow monitor the quality of this key. 

The most basic way to do so is to check the fraction of pulses causing a click in Bob's detection system and the fraction of errors among all his measurement results and compare them with their respective expected values, which can be calculated by Alice and Bob basing on the parameters of their setup. The possibility for them to perform such a task during the key generation process imposes some constraints on Eve's activities because she has to make sure that the expected values of these two quantities will be indeed recreated. Otherwise the legitimate participants of a given QKD protocol would immediately get suspicious of Eve's presence and deem the generated key insecure. The limitations on the eavesdropper's actions may be even greater if Alice and Bob use the so-called decoy-pulse method \cite{Hwang03,Wang05,*Lo05} during the QKD process. In this case, while Alice randomly changes the intensity of light emitted by her realistic source, Bob checks fractions of pulses causing a click in his detection system and fractions of errors among all his measurement results for each of the different intensities separately. In the most favorable situation, from the point of view of Alice and Bob, decoy-pulse method may totally prevent Eve from manipulating the statistics of photons reaching Bob, which she could otherwise safely do \emph{e.g.}\,by sending some of them through a lossless quantum channel and blocking the others.

Another useful way for monitoring the security of a given QKD scheme, based on checking of the violation of Bell inequalities, can be applied by Alice and Bob in the case of the so-called entanglement-based protocols \cite{Ekert91,Jennewein00,*Naik00,*Tittel00,Acin06}. In this situation a central source of light, independent from the trusted parties, produces entangled pairs of photons. One photon from each of these pairs is subsequently sent to Alice, while the other one is sent to Bob. Due to the quantum correlations between the two parts of a given entangled state, the results of the measurements performed by Alice and Bob on their respective parts of the state are correlated with each other. A good way to monitor such correlations (and thus the quality of the generated key) is to check if these results violate Bell inequalities and by how much they do so. Such an idea for \emph{a priori} verifying suitability of a given entanglement-based DV QKD protocol for secure quantum communication was proposed by Ekert in 1991 along with his famous E91 protocol \cite{Ekert91} and later became the essential part of the so-called device-independent (DI) \cite{Barrett05,Acin07} and measurement-device-independent (MDI) QKD schemes \cite{Lo12,Liu13,*Ferreira13,Pirandola15}.

Although DI and MDI QKD protocols allow Alice and Bob to release assumptions about trusted nature of their devices, their implementation is quite challenging due to the necessity of performing long-distance single-photon interference, which remains a demanding task even in the laboratory. Due to this limitation, the conventional prepare-and-measure QKD protocols, utilizing single photons sent from Alice to Bob, remain more suitable for practical implementations. In this case the verification whether the resource of security is preserved after the channel can be made by monitoring quantum features of photons arriving at Bob's detection system. It can be done by performing the measurement of the so-called autocorrelation function.  This procedure has been used for a long time as a useful tool for checking the property of quantum nonclassicality \cite{Glauber63,*Grangier86,*Hong87,*Filip13} and, very recently, quantum non-Gaussianity \cite{Filip11,*Jezek11} of realistic single-photon states. The latter one of these two properties excludes every possibility to describe experimental data using any mixture of Gaussian states \cite{Genoni07,*Genoni10,*Weedbrook12}. It represents a higher threshold for nonclassical features of light, which is much challenging to overcome. Simultaneously, it is very robust as has been predicted and experimentally verified for realistic single photon sources \cite{Lachman13,Straka14}.

In this paper we compare quantum nonclassicality and non-Gaussianity criteria with security conditions for DV QKD protocols over noisy quantum channels. It is well known, that while QKD systems can in principle tolerate any amount of loss in the quantum channel, the noise present in the channel limits the performance and secure distance of the protocols. In order to study security of DV QKD in the noisy channels and compare it to the quantum non-Gaussianity and nonclassicality criteria, we introduce three typical physical realizations of channel noise, which differ by the noise statistics and coupling to the signal. A similar approach was recently used to model a quantum channel in the study of microwave quantum communication \cite{Xiang16}. However, we assume that an eavesdropper is able to purify the channel noise, i.e. has the quantum channel under full control, and is able to perform optimal collective attacks. Without the loss of generality we perform our investigations for the case of Alice and Bob using BB84 protocol \cite{Bennett84} based on polarization encoding. The results of our analysis show that nonclassicality of light reaching Bob's measurement system can be always treated as a necessary indication of further QKD security in each of the studied cases. On the other hand, non-Gaussianity of this light becomes a sufficient and simple indication of possible further QKD security in the case when the probability of registering a signal photon by Bob is small but considerably larger than the probability of registering a dark count in modern realistic detectors. Moreover it is also sufficient for the schemes with arbitrarily high photon production and transmission efficiency if the polarization of the  signal photons is not being disturbed during their propagation between the legitimate participants of a QKD protocol. On the other hand, if such a disturbance may occur the probability for Bob to register an error during his polarization measurement of a signal state influences both the maximal and the minimal transmittance of the quantum channel above and below which non-Gaussianity of light arriving at Bob's detection system stops being a relevant indication of possible QKD security. In any case, the actual security of a QKD protocol can be verified only upon full implementation of the protocol, while non-Gaussianity is suggested only as a relevant and sufficient pre-check whether a particular set-up including a noisy quantum channel can in principle be used for QKD.

The paper is organized as follows. In Sec.\,\ref{Sec:General} we briefly discuss how the security of BB84 protocol can be evaluated in realistic cases and introduce some recently found criteria for nonclassicality and non-Gaussianity of light emitted by realistic single-photon sources. Next, in Sec.\,\ref{Sec:Models} we give a detailed description of all the physical realizations of noisy channels that we consider by introducing three specific DV QKD models. Then we apply the general formulas introduced in Sec.\,\ref{Sec:General} to find the three aforementioned types of criteria for all the models in various situations. The derivation of analytical versions of these criteria and the expressions for the minimal secure value of the transmittance of the channel connecting Alice and Bob in each of these cases can be found in Sec.\,\ref{Sec:Approximations}. Finally, in Sec.\,\ref{Sec:Summary} we summarize our work.

\section{General assumptions and formulas}
\label{Sec:General}

\subsection{QKD security}

In order to assess the security of all the models for the DV QKD schemes considered in this paper, we focus on finding lower bounds on the quantity called secret fraction \cite{Scarani09}, which describes the amount of common secret key that can be distilled by Alice and Bob from their individual versions of the raw key through the procedures of error correction and privacy amplification, per one bit of the raw key. Using the quantum generalization of Csisz\'{a}r-K\"{o}rner theorem \cite{Csiszar78}, this bound can be calculated by using the following formula \cite{Devetak05}:
\begin{equation}
\Delta I=\max[0,I_{AB}-\min\left\{I_{EA},I_{EB}\right\}],
\label{eq:DeltaImain}
\end{equation}
where $I_{AB}$ is the mutual information between Alice's and Bob's raw keys and $I_{EA}$ ($I_{EB}$) represent the upper bound on the amount of information Eve can get on Alice's (Bob's) version of the raw key thanks to her attacks. 

If Alice's source emits only genuine single-photon pulses, Eve can maximize her information by performing general collective attacks on them. For BB84 protocol the amount of information she can get in this way can be upper-bounded by $I_{EA}=I_{EB}=H(Q)$ \cite{Kraus05,*Renner05}, where $H(Q)$ is Shannon entropy and $Q$ represents the so-called quantum bit error rate (QBER) estimated by Alice and Bob in the raw key. On the other hand, in more realistic situation when Alice's source emits also multiphoton pulses with nonzero probability, Eve's best strategy is to: i) perform so-called photon-number-splitting (PNS) attacks on all multiphoton pulses -- which in the case of BB84 protocol can give her all the information encoded in them, ii) make sure that Bob registers all of these pulses by forwarding them to him through some lossless quantum channel and iii) perform general collective attacks on such an amount of single-photon pulses, which is required for Eve to recreate the expected level of QBER in Bob's version of the key. If the number of pulses Eve is able to attack using this strategy is lower than the amount of clicks expected by Bob in his detection system during the whole key generation process, she also has to forward to him some of the remaining single-photon pulses without attacking them. The rest of the pulses may be blocked by her. 

As we have already stated in the introduction to this article, the possibility of applying this kind of strategy by Eve may be restricted by Alice and Bob if they use decoy-pulse method. However, we do not consider this method in our work. It would be useless in the cases of DV QKD models introduced in Sec.\,\ref{Sec:bath} and \ref{Sec:before} since both of them assume that Alice's source never emits multiphoton pulses. On the other hand applying decoy-pulse method to the model analyzed in Sec.\,\ref{Sec:SPDC} would simply reduce this scheme to the one analyzed in Sec.\,\ref{Sec:bath} (in the best case scenario of infinitely many decoy intensities used by Alice). We also do not consider here the so-called preprocessing procedure \cite{Kraus05,*Renner05}, which could be utilized by Alice and Bob to increase the security of the key by deliberately adding some noise to it. This procedure would not affect the results of our analysis in significant way and would not change the general conclusions which can be derived from them. 

In the situation when decoy-pulse method and the preprocessing procedure are not used by the trusted parties, the mutual information between Alice and Bob on the generated key can be written as $I_{AB}=1-H(Q)$ and the lower bound for $\Delta I$ can be expressed as \cite{Gottesman04}
\begin{equation}
\Delta I=\max\left[0,y-H(Q)-yH\left(\frac{Q}{y}\right)\right],
\label{eq:DeltaIreal}
\end{equation}
where $y$ is the fraction of genuine single-photon pulses among all the pulses registered by Bob's detectors. In the case when Alice's source never emits multiphoton pulses, the above expression obviously simplifies to 
\begin{equation}
\Delta I=\max\left[0,1-2H(Q)\right].
\label{eq:DeltaIideal}
\end{equation}

\subsection{Nonclassicality and non-Gaussianity measurements}
\label{Sec:NCNG}

\begin{figure}
\centering
\includegraphics[width=0.6\linewidth]{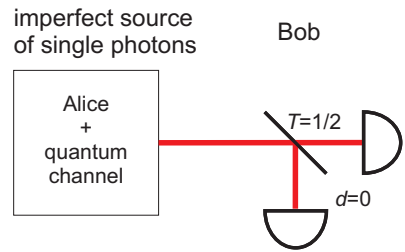}
\caption{(color online) A setup for the detection of nonclassicality and non-Gaussianity of light arriving at Bob's station, applied for all of the DV QKD models introduced in Sec.\,\ref{Sec:Models}.}
\label{fig:NCNGsetup}
\end{figure}

In what follows we compare the requirements for QKD security, calculated according to the general formula discussed above, with the criteria for nonclassicality and non-Gaussianity of light arriving at Bob's detection system. In order to find such criteria for the DV QKD models introduced further in Sec.\,\ref{Sec:Models} we assume that in the case of every particular model Bob can switch his QKD detection system to the standard setup for autocorrelation function measurement, consisting of 50:50 beam-splitter and two single-photon detectors, pictured in Fig.\,\ref{fig:NCNGsetup}. The rest of a given QKD setup, \emph{i.e.} Alice's source and the quantum channel along with any possible external sources of noise, coupling to the signal during its propagation from Alice to Bob, can be treated by him as an imperfect source of single photons. We focus on analyzing the situation in which Bob uses ideal on-off binary detectors with unity efficiency and no dark counts for his verification of nonclassicality and non-Gaussianity of the incomning light. Thanks to this assumption it is possible to check how QKD security can be related to the genuine nonclassicality and non-Gaussianity of light arriving at Bob's station, undisturbed by any potential imperfections in the process of autocorrelation function measurement. However, in the Appendix \ref{Sec:RealDetection} we also briefly discuss more practical scenario, in which Bob utilizes exactly the same realistic single-photon detectors with non-zero dark count probability for the generation of secure key and for the verification of nonclassicality and non-Gaussianity of light arriving at his station.

In our study we use the results of the analysis of nonclassicality and non-Gaussianity of light emitted by an imperfect source of single photons performed in \cite{Lachman13}. According to this paper, if we denote the probability of registering a single detection event in one of the detectors pictured in Fig.\,\ref{fig:NCNGsetup} by $P_{S}$ and the probability of a coincidence of detection events by $P_C$, the boundary for nonclassicality of light emitted by a given source can be found by solving the equation
\begin{equation}
P_S=2\left(\sqrt{P_C}-P_C\right).
\label{eq:nonC}
\end{equation}
Finding a similar analytical expression for the non-Gaussianity boundary turns out to be impossible, but for a given value of $P_S$ the maximal value of $P_C$ for which light emitted from a particular source is non-Gaussian can be found by numerically solving the following pair of equations
\begin{equation}
\left\{\begin{array}{l}
1-\frac{P_S}{2}-P_C=\frac{4\sqrt{V}\exp\left[-n/(6+2V)\right]}{\sqrt{(3V+1)(3+V})}\\
1-P_S-P_C=\frac{2\sqrt{V}\exp\left[-n/(2+2V)\right]}{V+1}
\end{array},\right.
\label{eq:NGfull}
\end{equation}
where
\begin{equation}
n=\frac{(1-V^2)(V+3)}{V(3V+1)}
\end{equation}
and maximizing the solution for $P_C$ on $V\in(0,1)$. Quantum non-Gaussianity threshold given by (\ref{eq:NGfull}) represents much more strict condition for realistic single photon states \cite{Jezek11} than quantum nonclassicality threshold (\ref{eq:nonC}). This relation be seen \emph{e.g.}\,in the fact that attenuation of a quantum channel does not break nonclassicality of light emitted by realistic single photon sources, contrary to its non-Gaussianity \cite{Straka14}.

As we will see in Sec.\,\ref{Sec:Models}, by measuring non-Gaussianity of light arriving at Bob's station in the way described in the previous paragraph, the legitimate participants of a given DV QKD protocol may be able to verify suitability of their prepare-and-measure type of the scheme for the generation of secure cryptographic key. It is worth noting here that since the polarization of the photons sent by Alice does not have any effect on the results of autocorrelation function measurement performed by Bob, this verification can be done at any stage of the protocol. Furthermore, Bob does not have to inform Alice about the precise time in which he decides to switch his measurement setup to the one presented in Fig.\,\ref{fig:NCNGsetup}. Thus Eve is also unable to get such information. This means that she cannot  mislead Alice and Bob by acting differently during the stages of non-Gaussianity verification and key generation.

\section{Analysis of specific DV QKD models}
\label{Sec:Models}

In our study we assume that Eve is able to perform general collective attacks in a noisy channel, which is fully under her control. However, the statistics of noise and its coupling to the signal may differ. Therefore, in this section we introduce and analyze three specific physical realizations of the noisy channels. In all the three cases we assume first of all that the polarization detection system used by Bob always consists of a single polarization beam-splitter and two binary on-off detectors with perfect detection efficiency and dark count probability per gate $d\ll 1$. Secondly, for every one of the analyzed schemes we consider the possibility of rotating the polarization of a given signal photon during its propagation from Alice to Bob. It is modelled by applying the following transformation to the state $|\psi\rangle$ emitted by Alice:
\begin{equation}
|\psi\rangle\langle\psi|\rightarrow(1-e)|\psi\rangle\langle\psi|+e\frac{\hat{\mathds{1}}}{2}.
\label{eq:PolarizationSwitch}
\end{equation}
This kind of the so-called depolarizing channel represents the influence of a completely unpolarized polarization noise and is typically the most detrimental polarization quantum channel. Therefore, we are clearly describing the worst case scenario here. If a given polarization channel is different in reality, an improvement can be realized. Further we describe and analyze each of the DV QKD models in details.

\subsection{Model with thermal bath}
\label{Sec:bath}

\begin{figure}
\centering
\includegraphics[width=1.0\linewidth]{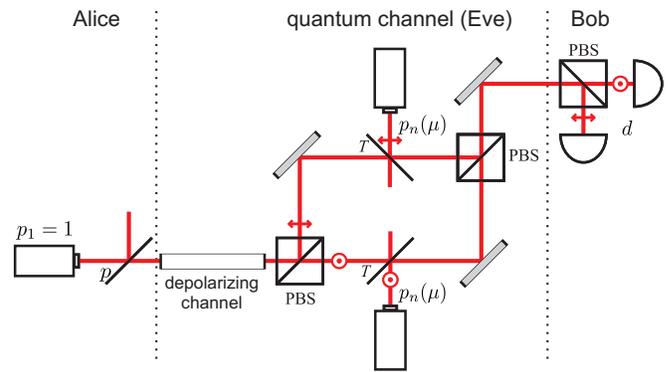}
\caption{(color online) The model for realistic DV QKD scheme with signal coupling to the channel noise in the form of the thermal bath.}
\label{fig:thermalbath}
\end{figure}

The model for the DV QKD scheme with the signal coupling to the noise in the form of the thermal bath inside the quantum channel, typical for continuous-variable quantum communication \cite{Lasota17}, is presented in Fig.\,\ref{fig:thermalbath}. In this configuration we assume that Alice's source never emits multiphoton pulses. It sends to Bob only one-photon and empty pulses with probabilities $p$ and $1-p$ respectively. The multimode channel connecting Alice and Bob is both lossy and noisy, and we describe its influence on a given signal state by coupling it to a thermal bath on a beam-splitter with transmittance $T$. This situation can be properly modeled by replacing the aforementioned channel by two single-mode channels transmitting orthogonal polarization modes, to which two independent and identically distributed sources of noise photons are being coupled, with the same polarizations respectively. We assume here that the statistics of photons generated by these sources are thermal, with mean numbers of photons emitted per pulse given by $\mu$. We denote the probability of generating $n$ noise photons by them in a single pulse by $p_n(\mu)$.

According to the formula (\ref{eq:DeltaIideal}), in order to assess the security of the QKD scheme presented in Fig.\,\ref{fig:thermalbath}, we need to express QBER in terms of the parameters describing the setup. Before we do this, it is useful to realize that since this configuration is totally symmetric in polarizations, it is not necessary to consider separately all of the cases when Alice's source sends differently polarized signal photons to Bob. It is sufficient to consider only a single case in which the polarization chosen by Alice and the detector to which the signal photon should go if Bob chose the right basis for his measurement are both called \emph{right}, while the opposite polarization and the other detector are called \emph{wrong}. Now by $p_+(k,l)$ let us denote the probability that the signal photon will be emitted by Alice and successfully transmitted through the channel into one of Bob's detectors and at the same time $k$ noise photons will arrive at the right detector, while $l$ noise photons will arrive at the wrong detector. Similarly, by $p_-(k,l)$ we will denote the probability that the signal photon will not reach Bob's detection system in a given attempt but nevertheless $k$ and $l$ noise photons will arrive at the right and wrong detectors respectively. These two quantities are equal to
\begin{equation}
p_+(k,l)=pT\pi_k(T)\pi_l(T)
\label{eq:pplus}
\end{equation}
and
\begin{equation}
p_-(k,l)=(1-pT)\pi_k(T)\pi_l(T),
\label{eq:pminus}
\end{equation}
where
\begin{equation}
\pi_k(T)=\sum_{n=k}^\infty p_n(\mu){n \choose k} (1-T)^kT^{n-k}.
\label{eq:pi}
\end{equation}

We assume here that Bob's detectors are not capable of resolving the number of incoming photons, so Alice and Bob automatically have to accept every event in which exactly one of these detectors clicks, no matter how many photons actually reached it. Since from (\ref{eq:pplus}) and (\ref{eq:pminus}) we have $p_\pm(k,l)=p_\pm(l,k)$ the expected probability of accepting a given event by Alice and Bob can be expressed as
\begin{equation}
p_{exp}^{(I)}=\sum_{k=0}^\infty p_+(k,0)+2\sum_{k=1}^\infty p_-(k,0)+2dp_-(0,0).
\label{eq:pexpsimple}
\end{equation}

An error in Bob's version of the key appears in the following cases: i) with probability $e/2$ -- when the signal photon survives the transmission through the channel and the noise photons arrive only at the same detector as the signal photon, ii) with $50\%$ probability -- when the only surviving photons are the noise photons traveling into one of the detectors and iii) with $50\%$ probability -- when there are no real photons arriving at the detectors and the click in one of them is caused by a dark count. This means that QBER expected by Alice and Bob is given by
\begin{equation}
Q^{(I)}=\frac{\frac{e}{2}\sum_{k=0}^\infty p_+(k,0)+\sum_{k=1}^\infty p_-(k,0)+dp_-(0,0)}{p_{exp}^{(I)}}.
\label{eq:Q}
\end{equation}

Having the formula for QBER derived above, we now check the security of the scheme illustrated in Fig.\,\ref{fig:thermalbath} for different values of setup parameters. However, if we also want to compare the security criterion with the criteria for nonclassicality and non-Gaussianity of light arriving at Bob's side, we need to find similar expressions for $P_S$ and $P_C$, defined in Sec.\,II.B, in the situation when Bob's detection system in Fig.\,\ref{fig:thermalbath} is replaced by the standard setup for the autocorrelation function measurement, pictured in Fig.\,\ref{fig:NCNGsetup}. Assuming that the detectors in this setup are perfect, these two quantities can be expressed in terms of probabilities $p_\pm(k,l)$ as follows:
\begin{equation}
P_S^{(I)}=\sum_{k,l=0}^\infty\frac{ p_+(k,l)+2p_-(k,l)}{2^{k+l}}-2p_-(0,0),
\label{eq:ps1}
\end{equation}
\begin{equation}
P_C^{(I)}=1-P_S^{(I)}-p_-(0,0).
\label{eq:pc1}
\end{equation}

\begin{figure}
\centering
\includegraphics[width=1.0\linewidth]{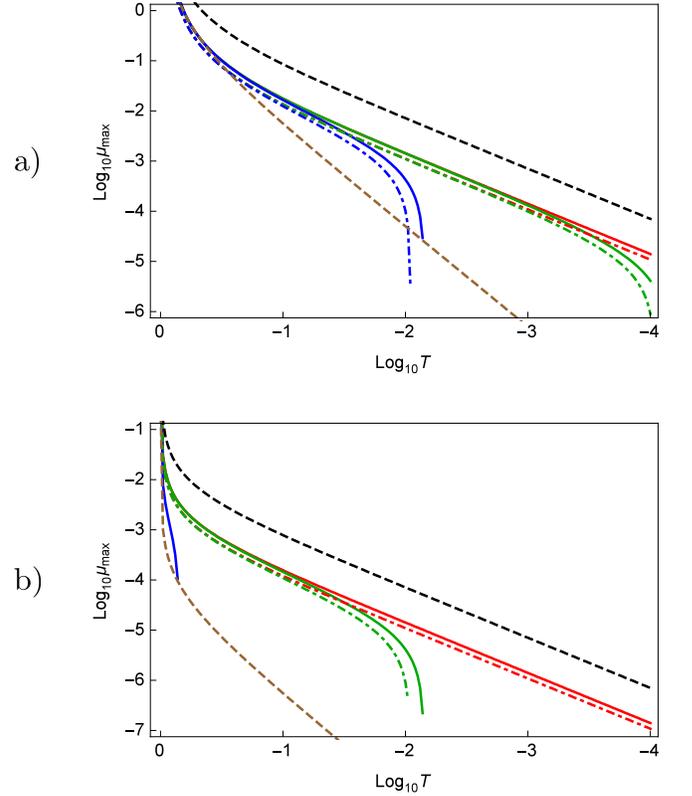}
\caption{(color online) Comparison between the lower bounds for the security of BB84 protocol realized using the scheme presented in Fig.\,\ref{fig:thermalbath} for a) $p=1$ and b) $p=0.01$ for two different values of the parameter $e$: $e=0$ (solid lines) and $e=0.05$ (dot-dashed lines), calculated numerically for the cases of $d=0$ (red, uppermost lines), $d=10^{-5}$ (green, middlemost lines) and $d=10^{-3}$ (blue, lowermost lines). The calculations were performed using the formula (\ref{eq:DeltaIideal}), with $Q$ given by the expression (\ref{eq:Q}). With black (upper) and brown (lower) dashed lines respectively we denoted the corresponding nonclassicality and non-Gaussianity criteria for the light arriving at Bob's detection system, calculated according to the equations (\ref{eq:nonC}) and (\ref{eq:NGfull}), with the quantities $P_S$ and $P_C$ given by the expressions (\ref{eq:ps1}) and (\ref{eq:pc1}).}
\label{fig:NCNGthermalbath1}
\end{figure}

In Fig.\,\ref{fig:NCNGthermalbath1} we show the dependency of the maximal power of the source of noise allowing for the generation of secure cryptographic key on the transmittance of the channel connecting Alice and Bob, calculated numerically for a few different values of the parameters $d$, $e$ and $p$ in the case of the QKD model analyzed above. In the same figure we also compare these security criteria with the similar criteria for nonclassicality and non-Gaussianity of light arriving at Bob's detection system. From this picture we can conclude that when $d\ll pT\ll 1$ non-Gaussianity criterion appears to be a sufficient indication of further QKD security for every realistic value of $e$. It means that once a quantum channel complies with the considered noise model, the observation of non-Gaussianity at the receiver station is fully sufficient for successful implementation of DV QKD protocol if only the error parameters $d$ and $e$ are negligibly small. Importantly, this conclusion remains valid for any $0<p\leq1$. Therefore, the indicator is a relevant {\em partial} sufficient security condition addressing output quantum statistics of single-mode light. It is very different to the quantum nonclassicality of light, which can be seen only as a necessary indicator of potential possibility to achieve the security (see Fig.\,\ref{fig:NCNGthermalbath1}).
However, the security of a particular realization of the DV QKD protocol can not be fully proven merely by the observation of non-Gaussianity. It can be verified only by estimating all the relevant parameters of the protocol, which define its security (\emph{i.e.} the level of QBER in the considered case).

\subsection{Model with noise coupling to the signal before the quantum channel}
\label{Sec:before}

\begin{figure}
\centering
\includegraphics[width=1.0\linewidth]{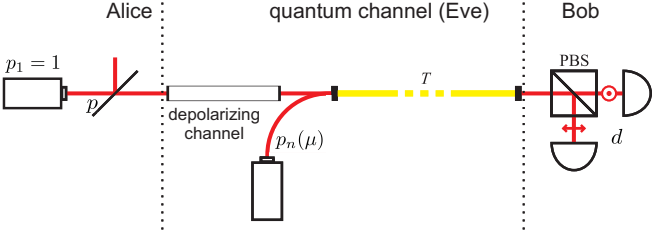}
\caption{(color online) The model for realistic DV QKD scheme with noise generated by an external source of light coupling to the signal before the quantum channel.}
\label{fig:noisebefore}
\end{figure}

The model for DV QKD scheme with noise coupling to the signal before the quantum channel is presented in Fig.\,\ref{fig:noisebefore}. Here our assumptions on the signal source owned by Alice are exactly the same as for the model with thermal bath analyzed previously and the channel connecting Alice and Bob is once again lossy with transmittance $T$. The difference between the schemes illustrated in Fig.\,\ref{fig:thermalbath} and Fig.\,\ref{fig:noisebefore} lies in the type of noise, coming from the external source of light (independent from Alice and Bob), which is added to the signal during its propagation between the trusted parties. This time the noise photons couple to the signal before the channel, meaning that they are being attenuated inside of it in the same way as the photons carrying the information about the key. Our basic assumption here is that the probability distribution $p_n(\mu)$ of the number of photons generated by the source of noise per pulse is of the thermal type. However, for comparison in the Appendix \ref{Sec:ThermalPoisson} we also consider the situation when it is given by Poisson statistics. Furthermore, we assume here that all of the noise photons in a given pulse are emitted in the same polarization state $|\psi\rangle$, which is a random superposition of the signal state $|A\rangle$ sent by Alice, to which this particular noise photons are coupled, and the state $|A^\perp\rangle$ which is orhogonal to it. In other words
\begin{equation}
|\psi\rangle=\sqrt{x}|A\rangle+e^{i\phi}\sqrt{1-x}|A^\perp\rangle,
\label{eq:randompolar}
\end{equation}
where $x$ can take all the values between $0$ and $1$ with uniformly distributed probability and $\phi\in[0,2\pi)$.

Since in the case of Alice and Bob using DV QKD scheme presented in Fig.\,\ref{fig:noisebefore} we again assume that all the double clicks in Bob's detection system are discarded from the key automatically, an important quantity for our security analysis of this setup configuration is the probability that all of the photons belonging to a particular $j$-photon noise pulse would go to the same detector. For photons in a given state of the form (\ref{eq:randompolar}) this probability can be calculated as
\begin{equation}
f(x)=x^j+(1-x)^j.
\end{equation}
It can be shown that the mean value of this quantity, randomized over every possible value of $x$ is equal to $2/(j+1)$.

To assess the security of the scheme presented in Fig.\,\ref{fig:noisebefore}, we again use the formula (\ref{eq:DeltaIideal}). In order to express the value of QBER in terms of the parameters describing this setup configuration let us denote the probability that only the signal photon (the noise photons) would arrive at Bob's measurement system and cause a click in one of his detectors by $p_{exp}^{s}$ ($p_{exp}^{n}$). Furthermore we  use $p_{exp}^{n,s}$ to describe the probability that both signal photon and at least one noise photon reach Bob's measurement system and cause a single click in one of his detectors, and $p_{exp}^{dc}$ for the case when no real photons manage to get through the quantum channel but there is a click on Bob's side in the form of a dark count. These four quantities can be calculated in the following way:
\begin{equation}
p_{exp}^{s}=pT\sum_{i=0}^\infty p_i(\mu)(1-T)^i,
\label{eq:pexps}
\end{equation}
\begin{equation}
p_{exp}^{n}=2(1-pT)\sum_{i=1}^\infty p_i(\mu)r_i(T),
\end{equation}
\begin{equation}
p_{exp}^{n,s}=pT\sum_{i=1}^\infty p_i(\mu)r_i(T)
\end{equation}
and
\begin{equation}
p_{exp}^{dc}=2d(1-pT)\sum_{i=0}^\infty p_i(\mu)(1-T)^i,
\end{equation}
where 
\begin{equation}
r_i(T)=\sum_{j=1}^{i}\frac{{i\choose j}T^j(1-T)^{i-j}}{j+1}
\end{equation}
is the probability that an $i$-photon pulse emitted by the source of noise will be successfully transmitted to Bob's detection system and all of the surviving photons from this pulse will go to the same detector. By using the above quantities QBER can be expressed in the following way:
\begin{equation}
Q^{(II)}=\frac{e\left(p_{exp}^{s}+p_{exp}^{n,s}\right)+p_{exp}^{n}+p_{exp}^{dc}}{2p_{exp}^{(II)}},
\label{eq:qber2}
\end{equation}
where 
\begin{equation}
p_{exp}^{(II)}=p_{exp}^{s}+p_{exp}^{n,s}+p_{exp}^{n}+p_{exp}^{dc}.
\end{equation} 

On the other hand, the probabilities $P_S$ and $P_C$, needed for the calculation of nonclassicality and non-Gaussianity criteria for light arriving at Bob's side, can be written as
\begin{equation}
P_S^{(II)}=p_{exp}^{s}+\tilde{p}_{exp}^{n,s}+\tilde{p}_{exp}^{n},
\label{eq:ps2}
\end{equation}
and
\begin{equation}
P_C^{(II)}=1-P_S^{(II)}-(1-pT)\sum_{i=0}^\infty p_i(\mu)(1-T)^i
\label{eq:pc2}
\end{equation}
where
\begin{equation}
\tilde{p}_{exp}^{n}=2(1-pT)\sum_{i=1}^\infty p_i(\mu)s_i(T),
\end{equation}
\begin{equation}
\tilde{p}_{exp}^{n,s}=pT\sum_{i=1}^\infty p_i(\mu)s_i(T)
\end{equation}
and
\begin{equation}
s_i(T)=\sum_{j=1}^{i}\frac{{i\choose j}T^j(1-T)^{i-j}}{2^j}
\label{eq:siT}
\end{equation}
is the probability that at least one out of $i$ photons emitted by the source of noise in a given pulse would survive its travel through the quantum channel and cause a click in a particular one of the two Bob's detectors in the detection system for the measurement of autocorrelation function, illustrated in Fig.\,\ref{fig:NCNGsetup}.

\begin{figure}
\centering
\includegraphics[width=1.0\linewidth]{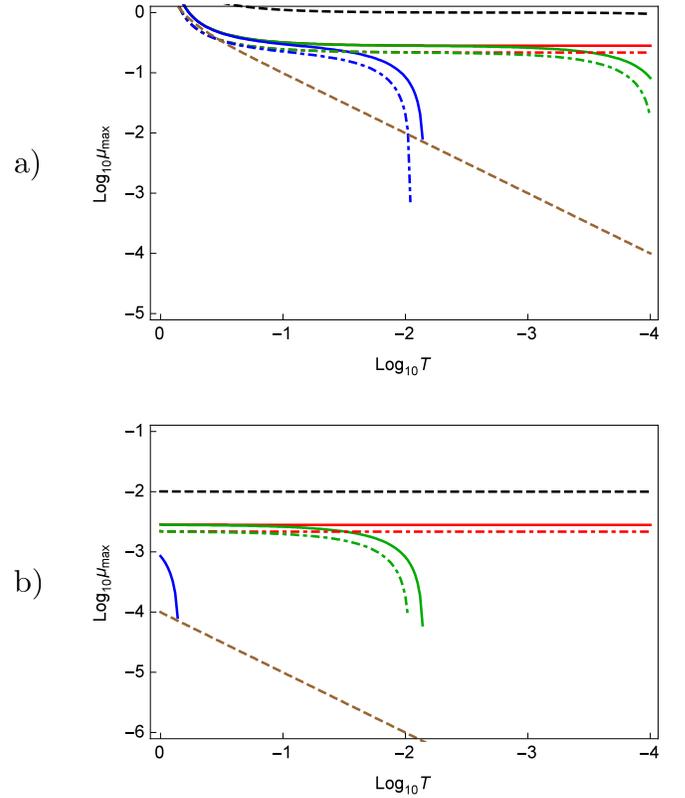}
\caption{(color online) Comparison between the lower bounds for the security of BB84 protocol realized using the scheme presented in Fig.\,\ref{fig:noisebefore} and nonclassicality and non-Gaussianity criteria for light arriving at Bob's side in this case, found in the situation when the source of noise has thermal statistics of the number of photons emitted per pulse.  The calculations of QKD security were performed using the formula (\ref{eq:DeltaIideal}), with $Q$ given by the expression (\ref{eq:qber2}). The nonclassicality and non-Gaussianity criteria were calculated according to the equations (\ref{eq:nonC}) and (\ref{eq:NGfull}), with the quantities $P_S$ and $P_C$ given by the expressions (\ref{eq:ps2}) and (\ref{eq:pc2}) respectively. The styles and colors of all the lines are exactly the same as in Fig.\,\ref{fig:NCNGthermalbath1}.}
\label{fig:NCNGnoisebeforechannel1}
\end{figure}

The dependency of the maximal power of the source of noise, for which it is possible to utilize the setup illustrated in Fig.\,\ref{fig:noisebefore} to generate secure cryptographic key by using BB84 protocol, on the transmittance of the channel connecting Alice and Bob, calculated numerically for a couple of values of $d$, $e$ and $p$ for the thermal type of statistics of the source of noise, is presented in Fig.\,\ref{fig:NCNGnoisebeforechannel1}. The resulting QKD security criteria are also compared there with the criteria for nonclassicality and non-Gaussianity of light arriving at Bob's detection system. The conclusions concerning the relationships between these different types of criteria that can be derived here are analogous to the ones regarding the model for the DV QKD scheme presented in Fig.\,\ref{fig:NCNGthermalbath1}, stated in the previous section. Nonclassicality of light arriving at Bob's detection system can be once again treated as a necessary indication of further QKD security in every realistic situation, while fulfillment of the non-Gaussianity criterion turns out to be sufficient for the security in most of the cases. It may be not sufficient only i) when the values of $p$ and $T$ are relatively high and $e>0$ at the same time or ii) when the probability of registering a real signal in Bob's measurement system is not much larger than the probability of registering a dark count. This remarkable similarity of the results of our analyses performed for the models illustrated in Fig.\,\ref{fig:thermalbath} and Fig.\,\ref{fig:noisebefore} suggests that the conclusions stated here may apply to every type of DV QKD scheme with perfect single-photon source and noisy quantum channel, regardless of the type of this noise.

\subsection{Model with thermal bath and SPDC-based signal source}
\label{Sec:SPDC}

\begin{figure}
\centering
\includegraphics[width=1.0\linewidth]{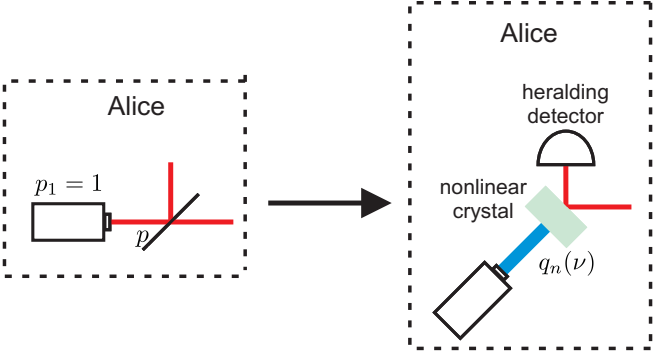}
\caption{(color online) The change of Alice's setup which should be made in the QKD schemes illustrated in Fig.\,\ref{fig:thermalbath} and Fig.\,\ref{fig:noisebefore} in order to perform their security analyses in the case when the key is being generated using realistic single-photon source based on SPDC process.}
\label{fig:SPDCchange}
\end{figure}

While the DV QKD schemes analyzed in the subsections \ref{Sec:bath} and \ref{Sec:before} deal with two different kinds of channel noise, both of them contain the same idealized model of the source of photons used by Alice. It is therefore reasonable to consider the influence of the possibility of generating multiphoton pulses by a more realistic signal source on the results presented in the previous subsections. In order to study this issue we once again consider the scheme illustrated in Fig.\,\ref{fig:thermalbath}, but with a substantial modification of Alice's part of the setup, presented in Fig.\,\ref{fig:SPDCchange}. As can be seen in this figure, we assume here that Alice owns heralded single-photon source based on the spontaneous parametric down-conversion (SPDC) process. In this type of source one photon from each pair created inside a nonlinear crystal, called idler, is always sent to a heralding detector, while the other one, called signal, is used to transmit the information about the key, but only if the heralding detector clicks. The photons for DV QKD applications are typically produced by a multimode SPDC process. In this case the probabilities of generating different numbers of photon pairs per one pulse of the pump laser can be approximated  by the Poisson statistics \cite{Helwig09}. If we denote the mean number of these pairs by $\nu$, we can express the state produced in the crystal as
\begin{equation}
\hat{\rho}=\sum_{n=0}^\infty e^{-\nu}\frac{\nu^n}{n!}|n\rangle_\mathrm{S}\langle n|\otimes|n\rangle_\mathrm{I}\langle n|,
\end{equation} 
where the subscripts $\mathrm{S}$ and $\mathrm{I}$ denote the Fock states belonging to the Hilbert spaces of signal and idler photons respectively. 

After emerging from the crystal idler photons are directed to the heralding detector. For simplicity we assume here that this device is ideal. It means that it never registers dark counts and successfully heralds any non-empty signal pulse emitted by Alice's source. Thus, its action on the state $\hat{\rho}$ can be expressed by the operator
\begin{equation}
\hat{P}_\mathrm{herald}=\hat{\mathds{1}}_\mathrm{I}-|0\rangle_\mathrm{I}\langle 0|
\end{equation}
and the probability that an $i$-photon signal pulse will be generated during the SPDC process and subsequently accepted for the QKD reads
\begin{equation}
q_i(\nu)=\,_\mathrm{S}\langle i|\mathrm{Tr}_\mathrm{I}\left[\hat{\mathds{1}}_\mathrm{S}\otimes\hat{P}_\mathrm{herald}\hat{\rho}\right]|i\rangle_\mathrm{S}.
\end{equation}
Therefore we have
\begin{equation}
q_i(\nu)=\left\{
\begin{array}{l}
0,\ i=0\\
e^{-\nu}\frac{\nu^i}{i!},\ i\neq0
\end{array}\right. .
\end{equation}

\begin{figure}
\centering
\includegraphics[width=1.0\linewidth]{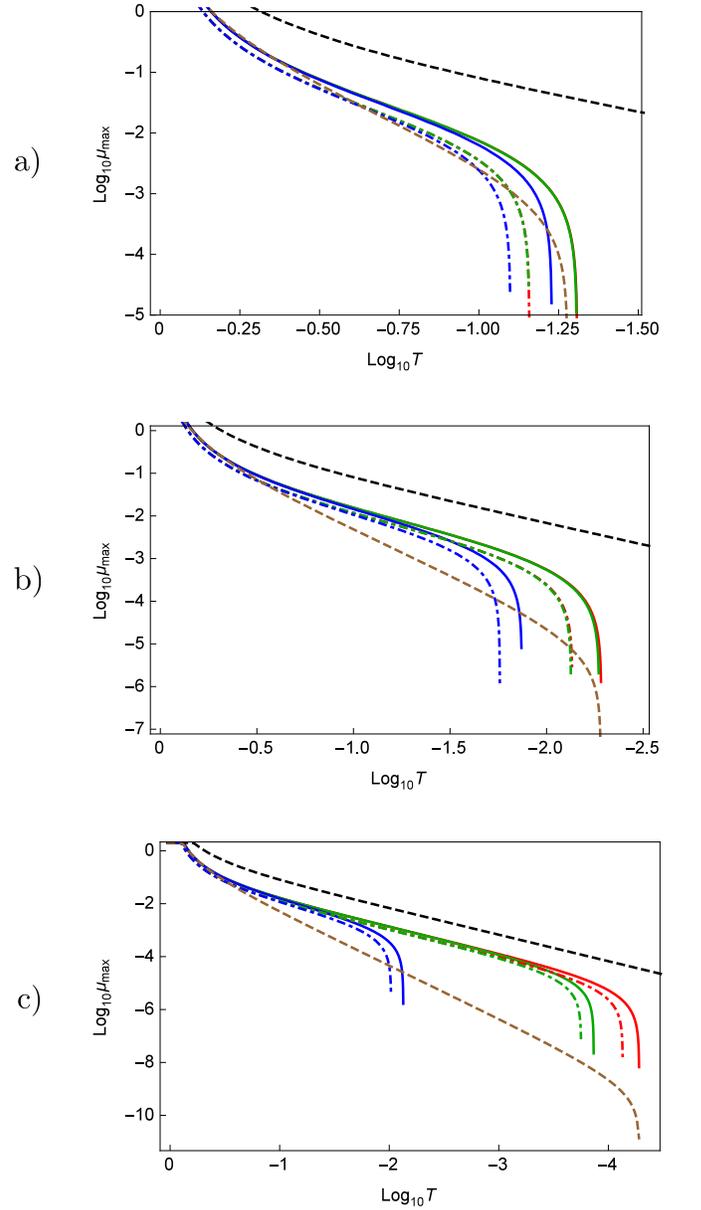}
\caption{(color online) Comparison between the numerically calculated lower bounds for the security of BB84 protocol realized using the scheme pictured in Fig.\,\ref{fig:thermalbath} with the modification presented in Fig.\,\ref{fig:SPDCchange} and nonclassicality and non-Gaussianity criteria for light arriving at Bob's side in this particular case. The three panels illustrate the plots made for the following values of the parameter $\nu$: a) $\nu=10^{-1}$, b) $\nu=10^{-2}$, c) $\nu=10^{-4}$. The calculations of QKD security were performed using the formula (\ref{eq:DeltaIreal}), with $y$ and $Q$ given by the expression (\ref{eq:y}) and (\ref{eq:qber3}) respectively. The nonclassicality and non-Gaussianity criteria were calculated according to the equations (\ref{eq:nonC}) and (\ref{eq:NGfull}), with the quantities $P_S$ and $P_C$ given by the expressions (\ref{eq:ps3}) and (\ref{eq:pc3}) respectively. The styles and colors of all the lines are exactly the same as in Fig.\,\ref{fig:NCNGthermalbath1}.}
\label{fig:NCNGSPDCgeneral1}
\end{figure}

Since Alice's source utilizing the down-conversion phenomenon emits multiphoton pulses with non-zero probability, we have to use the formula (\ref{eq:DeltaIreal}) if we want to properly analyze the security of the modified thermal bath DV QKD model. This means that beside QBER, we also have to find the formula for the fraction of clicks $y$ caused by genuine single-photon pulses among all the clicks registered by Bob. Before we do this, let us define the following functions:
\begin{equation}
P_+(k,l)=\sum_{i=0}^\infty q_i(\nu)t_i(T)\pi_k(T)\pi_l(T)
\label{eq:pplusmodif}
\end{equation} 
and
\begin{equation}
P_-(k,l)=\sum_{i=0}^\infty q_i(\nu)(1-t_i(T))\pi_k(T)\pi_l(T)
\label{eq:pminusmodif},
\end{equation}
where the quantity $\pi_k(T)$ is given by the expression (\ref{eq:pi}) and $t_i(T)$ denotes the probability that at least one photon emitted by Alice in a given \textit{i}-photon signal pulse will survive its travel to Bob's measurement system. This probability is equal to
\begin{equation}
t_i(T)=\sum_{j=1}^i{i \choose j}T^j(1-T)^{i-j}.
\end{equation}
Since in the worst case scenario Eve can forward to Bob all of the multiphoton pulses via lossless quantum channel, we have
\begin{equation}
y=\max\left[0,\frac{p_{exp}^{(III)}-p_{multi}}{p_{exp}^{(III)}}\right],
\label{eq:y}
\end{equation}
where
\begin{equation}
p_{exp}^{(III)}=\sum_{k=0}^\infty P_+(k,0)+2\sum_{k=1}^\infty P_-(k,0)+2dP_-(0,0)
\label{eq:pexp3}
\end{equation}
and
\begin{equation}
p_{multi}=\sum_{i=2}^\infty q_i(\nu).
\label{eq:pmulti}
\end{equation}
The QBER produced by the modified thermal bath DV QKD model can be written as
\begin{equation}
Q^{(III)}=\frac{\frac{e}{2}\sum_{k=0}^\infty P_+(k,0)+\sum_{k=1}^\infty P_-(k,0)+dP_-(0,0)}{p_{exp}^{(III)}}.
\label{eq:qber3}
\end{equation}
On the other hand, the quantities $P_S$ and $P_C$, needed for the evaluation of nonclassicality and non-Gaussianity criteria, are now equal to
\begin{equation}
P_S^{(III)}=\frac{\sum_{i,k,l=0}^\infty 2^{1-k-l}q_i(\nu)u_i(T)\pi_k(T)\pi_l(T)}{\sum_{i=0}^\infty q_i(\nu)}
\label{eq:ps3}
\end{equation}
and
\begin{equation}
P_C^{(III)}=1-P_S^{(III)}-\frac{\sum_{i=0}^\infty q_i(\nu)(1-T)^i\left(\pi_0(T)\right)^2}{\sum_{i=0}^\infty q_i(\nu)},
\label{eq:pc3}
\end{equation}
where
\begin{equation}
u_i(T)=\sum_{j=\max[0,1-k-l]}^i\frac{{i \choose j}T^j(1-T)^{i-j}}{2^j}.
\end{equation}

Utilizing the expressions derived above we numerically found the criteria for the security of the modified thermal bath DV QKD scheme considered in the present subsection, and for nonclassicality and non-Gaussianity of light arriving at Bob's measurement system in this particular case. Our calculations were performed for a few different values of the parameters $d$, $e$ and $\nu$. The resulting $\mu_{max}(T)$ functions are plotted in Fig.\,\ref{fig:NCNGSPDCgeneral1}. The conclusions that can be derived from this figure are once again very similar to the ones stated in Sec.\,\ref{Sec:bath}. The only substantial difference between the results obtained in the present analysis and the ones pictured in Fig.\ref{fig:NCNGthermalbath1} is the fact that when the process of key generation is plagued at the same time by both the channel noise and the multiphoton signal pulses, there always exist some minimal values of $T$ below which non-Gaussianity or QKD security cannot be obtained, even if $d=0$. However, as long as $e=0$ and $d\ll T$ the threshold value of $T$ required for non-Gaussianity of light arriving at Bob's detection system is approximately equal to the analogous value for QKD security (or even higher if $\nu$ is relatively high at the same time, as can be seen in Fig.\ref{fig:NCNGSPDCgeneral1} a)). Thus, in this situation non-Gaussianity criterion remains the sufficient indication of further QKD security for every $T\ll1$, similarly as in the model described in Sec.\,\ref{Sec:bath}.

It should be mentioned here that the results of the analysis presented in Fig.\,\ref{fig:NCNGSPDCgeneral1} would not change significantly if we assumed limited efficiency of the heralding detector in our model for SPDC-based signal source described above. Also the dark counts which could be registered by this measurement device in realistic situation would not affect our considerations unless a very low-efficient SPDC process was performed, with probability of emitting a non-empty pulse comparable with the probability of registering a dark count by the heralding detector.

\section{Analytical approximations of various criteria}
\label{Sec:Approximations}

\begin{figure}
\centering
\includegraphics[width=1.0\linewidth]{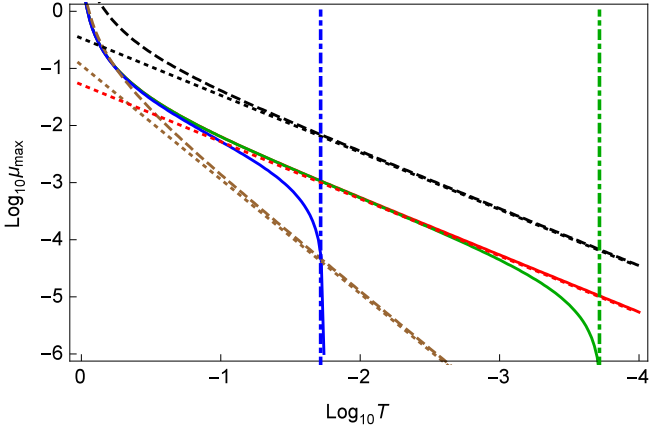}
\caption{(color online) Comparison between the numerically calculated nonclassicality (dashed black, upper line), and non-Gaussianity (dashed brown, lower line) criteria and the lower bounds for the security of BB84 protocol (solid lines) realized by using DV QKD scheme presented on Fig.\,\ref{fig:thermalbath} for $p=0.5$ and $e=0.05$ in the cases of $d=0$ (red, uppermost line), $d=10^{-5}$ (green, middlemost line) and $d=10^{-3}$ (blue, lowermost line), with the analogous analytical criteria (corresponding dotted lines)  calculated for $T\rightarrow 0$ and $d=0$, using formulas (\ref{eq:mumaxQKD}), (\ref{eq:mumaxNC}) and (\ref{eq:mumaxNG}). Additionally dot-dashed green (right) and blue (left) lines denote minimal secure values of $T$, analytically calculated for the cases of $d=10^{-5}$ and $d=10^{-3}$ respectively, using the formula (\ref{eq:Tmin}). The numerical calculations regarding QKD security, nonclassicality and non-Gaussianity criteria were performed using exactly the same formulas as in the case presented in Fig.\ref{fig:NCNGthermalbath1}.}
\label{fig:NCNGthermalbath3}
\end{figure}

While derivation of simple, analytical formulas for the security of the DV QKD models studied in this work and the nonclassicality and non-Gaussianity of light arriving at Bob's detection system is impossible in the general case, it can be done in some cases when $T\rightarrow 0$. Furthermore, for the case of $d\neq0$ it is possible to find approximate expression for the minimal transmittance of the channel connecting Alice and Bob for which a given QKD scheme can be secure. In this section we explain in details how to find the aforementioned approximations. Their quality can be assessed by analyzing Fig.\,\ref{fig:NCNGthermalbath3}, where all kinds of criteria calculated numerically for the DV QKD model with channel noise in the form of thermal bath, pictured in Fig.\,\ref{fig:thermalbath}, are plotted along with their respective analytical approximations. 

\subsection{QKD security criteria}
\label{Sec:MuQBER}

In order to derive analytical formulas for QKD security criteria in the case of $T\rightarrow 0$ for the two models analyzed in Sec.\,\ref{Sec:bath} and Sec.\,\ref{Sec:before} we have to simplify the formulas for their respective quantum bit error rates under the assumptions that $T\ll1$ and $d=0$ (if $d\neq0$ none of the schemes analyzed in this article is secure in the limit of $T\rightarrow 0$), and compare them with the threshold value of QBER, which for the case of BB84 protocol is approximately equal to $11\%$ \cite{Kraus05,*Renner05}. For the DV QKD model with thermal bath we find that
\begin{equation}
Q^{(I)}(T\rightarrow 0,d=0)\approx\frac{\frac{e}{2}pT+\frac{\mu}{\mu+1}}{pT+2\frac{\mu}{\mu+1}}.
\label{eq:Qapprox}
\end{equation}
Utilizing this result and noticing that for $T\ll1$ also the inequality $\mu_{max}(T)\ll1$ is always true, from the equation $Q^{(I)}=Q_{th}$ it is possible to get the following security criterion
\begin{equation}
\mu_{max}^{QKD,I}(T\rightarrow 0,d=0)\approx\frac{p\left(2Q_{th}-e\right)}{2(1-2Q_{th})}T.
\label{eq:mumaxQKD}
\end{equation}
On the other hand for the DV QKD model with noise coupling to the signal before the lossy quantum channel connecting Alice and Bob, described in Sec.\,\ref{Sec:before}, one has
\begin{equation}
Q^{(II)}(T\rightarrow 0,d=0)\approx\frac{(ep+\mu)}{2(p+\mu)}.
\label{eq:Qapprox2}
\end{equation}
Thus, the low-transmittance approximation of the function $\mu_{max}^{QKD,II}(T)$ takes the form
\begin{equation}
\mu_{max}^{QKD,II}(T\rightarrow 0,d=0)\approx\frac{p\left(2Q_{th}-e\right)}{1-2Q_{th}}.
\label{eq:mumaxQKD2}
\end{equation}
Finally, in the case of the modified thermal bath DV QKD model with SPDC-based signal source, analyzed in Sec.\,\ref{Sec:SPDC}, derivation of a simple analytical expression for the function $\mu_{max}^{QKD}(T)$ is possible only with the additional assumptions that $\nu\ll T$, which prevents multiphoton pulses to become a significant fraction of all the pulses registered by Bob during the key generation process. The resulting QKD security criterion reads
\begin{equation}
\mu_{max}^{QKD,III}(\nu\ll T\ll 1,d=0)\approx\frac{\left(2Q_{th}-e\right)}{2(1-2Q_{th})}T.
\label{eq:mumaxQKD3}
\end{equation}
As should be expected, for $p=1$ the expressions (\ref{eq:mumaxQKD}) and (\ref{eq:mumaxQKD3}) are identical to each other. The reason for this is the fact that in the limit of $\nu\rightarrow 0$ the SPDC-based signal source with perfect heralding detector becomes an ideal single-photon source with $100\%$ probability of emitting one photon in a given accepted signal pulse.

\subsection{Nonclassicality  and non-Gaussianity criteria}

In order to find approximate versions of the nonclassicality and non-Gaussianity criteria for all three DV QKD models introduced in Sec.\,\ref{Sec:Models} let us first denote  by $\omega_1$ ($\omega_{2+}$) the probability that there will be exactly one (more than one) photon arriving at Bob's side in a given pulse during the key generation process. It was shown in \cite{Lachman13} that if the condition $\omega_{2+}\ll \omega_1$ is fulfilled, these two quantities can be used to form much simpler nonclassicality and non-Gaussianity criteria, than the ones introduced in Sec.\,\ref{Sec:NCNG}. These simplified criteria can be written respectively as
\begin{equation}
\frac{\omega_1^2}{2}>\omega_{2+}
\label{eq:nonclass}
\end{equation}
and
\begin{equation}
\omega_1^3>\omega_{2+}.
\label{eq:nonGauss}
\end{equation}

In the case of DV QKD model with thermal bath, introduced in Sec.\,\ref{Sec:bath}, $\omega_1$ and $\omega_{2+}$ can be expressed by using the probabilities $p_\pm(k,l)$, defined by the formulas (\ref{eq:pplus}) and (\ref{eq:pminus}), as
\begin{equation}
\omega_1^{(I)}=p_+(0,0)+2p_-(1,0)
\label{eq:q1}
\end{equation}
and
\begin{equation}
\omega_{2+}^{(I)}=1-\omega_1^{(I)}-p_-(0,0).
\label{eq:q2plus}
\end{equation}
Since for $T\ll1$ and $\mu\ll1$ the relationship $\omega_{2+}^{(I)}\ll \omega_1^{(I)}$  is fulfilled, the inequalities (\ref{eq:nonclass}) and (\ref{eq:nonGauss}) can be indeed used to find the desired criteria in this situation. In the end they can be written in the following forms:
\begin{equation}
\mu_{max}^{NC,I}(T\rightarrow 0)\approx\frac{p}{\sqrt{2}}T
\label{eq:mumaxNC}
\end{equation}
for nonclassicality of the state reaching Bob's detection system and
\begin{equation}
\mu_{max}^{NG,I}(T\rightarrow 0)\approx\frac{p^2}{2}T^2
\label{eq:mumaxNG}
\end{equation}
for its non-Gaussianity.

To derive the analogous formulas for $\mu_{max}^{NC}(T\rightarrow 0)$ and $\mu_{max}^{NG}(T\rightarrow 0)$ for the DV QKD model analyzed in Sec.\,\ref{Sec:before} we can once again start with expressing the probabilities $\omega_1$ and $\omega_{2+}$ in terms of the parameters describing this particular scheme. We have
\begin{equation}
\omega_1^{(II)}=p_{exp}^s+(1-pT)\sum_{i=0}^\infty p_i(\mu)iT(1-T)^{(i-1)}
\label{eq:q1_2}
\end{equation}
and
\begin{equation}
\omega_{2+}^{(II)}=1-\omega^{(II)}_1-(1-pT)\sum_{i=0}^\infty p_i(\mu)(1-T)^i,
\label{eq:q2plus_2}
\end{equation}
where $p_{exp}^s$ has already been defined by the formula (\ref{eq:pexps}) in Sec.\,\ref{Sec:before}. For $T\ll 1$, when the statistics of the source of noise is thermal, these two quantities simplify to $\omega_1^{(II)}\approx(p+\mu)T$ and $\omega_{2+}^{(II)}\approx\mu(p+\mu)T^2$. If so, then from the criterion for nonclassicality (\ref{eq:nonclass}) we get
\begin{equation}
\mu_{max}^{NC,II}(T\rightarrow 0)\approx p.
\label{eq:mumaxNC2}
\end{equation}
Since for $T\rightarrow 0$ the value of the function $\mu_{max}^{NG,II}(T)$ also goes to zero (see Fig.\,\ref{fig:NCNGnoisebeforechannel1}), we can assume $\mu\ll p$ for the purpose of finding the simplest possible form of non-Gaussianity criterion for this QKD model. In the end it takes the following form:
\begin{equation}
\mu_{max}^{NG,II}(T\rightarrow 0)\approx p^2T.
\label{eq:mumaxNG2}
\end{equation}
As a side note it is worth mentioning here that if the source of noise in the scheme pictured in Fig.\,\ref{fig:noisebefore} had Poisson statistics, for $T\ll1$ one would get $\omega_{2+}^{(II)}\approx\mu(p+\mu/2)T^2$ and the same formula for $\omega_1^{(II)}$ as for the case of thermal statistics.  The non-Gaussianity criterion which could be derived in this situation would be the same as (\ref{eq:mumaxNG2}), while the criterion for nonclassicality would take the form $p>0$, which is of course always fulfilled.

In the case of the modified thermal bath QKD model with SPDC-based signal source, analyzed in Sec.\,\ref{Sec:SPDC}, the probabilities $\omega_1$ and $\omega_2$ are equal to
\begin{equation}
\omega_1^{(III)}=\frac{\sum_{i=0}^\infty q_i(\nu)(1-T)^{i-1}\Omega_i(T)}{\sum_{i=0}^\infty q_i(\nu)}
\label{eq:omega13}
\end{equation}
and
\begin{equation}
\omega_{2+}^{(III)}=1-\omega_1^{(III)}-\frac{\sum_{i=0}^\infty q_i(\nu)(1-T)^i\left(\pi_0(T)\right)^2}{\sum_{i=0}^\infty q_i(\nu)},
\label{eq:omega2plus3}
\end{equation}
where
\begin{equation}
\Omega_i(T)=iT\left(\pi_0(T)\right)^2+2(1-T)\pi_0(T)\pi_1(T).
\end{equation}
When the assumptions $\mu\ll 1$ and $\nu\ll T\ll 1$ are fulfilled, the expressions (\ref{eq:omega13}) and (\ref{eq:omega2plus3}) simplify to
\begin{equation}
\omega_{1}^{(III)}(\mu\ll 1, \nu\ll T\ll 1)\approx T+2\mu
\end{equation}
and
\begin{equation}
\omega_{2+}^{(III)}(\mu\ll 1, \nu\ll T\ll 1)\approx 2T\mu+3\mu^2
\end{equation}
respectively. In this case, using the inequalities (\ref{eq:q1}) and (\ref{eq:q2plus}), one can derive the following approximate criteria:
\begin{equation}
\mu_{max}^{NC,III}(\nu\ll T\ll 1)\approx\frac{T}{\sqrt{2}}
\label{eq:mumaxNC3}
\end{equation}
for nonclassicality and 
\begin{equation}
\mu_{max}^{NG,III}(\nu\ll T\ll 1)\approx\frac{T^2}{2}
\label{eq:mumaxNG3}
\end{equation}
for non-Gaussianity of light arriving at Bob's measurement system. As could be expected, for $p=1$ the analogous criteria (\ref{eq:mumaxNC}) and (\ref{eq:mumaxNG}) have the same form as (\ref{eq:mumaxNC3}) and (\ref{eq:mumaxNG3}) respectively.

\subsection{Minimal secure transmittance of the channel}

In this subsection for every one of the DV QKD models analyzed in this paper we find an approximate analytical formula for the minimal transmittance of the channel connecting Alice and Bob ($T_{min}$) for which it is possible to perform secure key generation process using BB84 protocol in the case when the probability of registering a dark count in Bob's detection system is non-zero. In order to perform this task for the schemes described in Sec.\,\ref{Sec:bath} and Sec.\,\ref{Sec:before} we once again simplify the expressions (\ref{eq:Q}) and (\ref{eq:qber2}) for $T\ll1$, like we did in Sec.\,\ref{Sec:MuQBER}, but this time with the assumption that $\mu\ll d$. In this situation we get
\begin{equation}
Q^{(I)}(T\ll 1,\mu\ll d)\approx\frac{\frac{e}{2}pT+d}{pT+2d}.
\end{equation}
and
\begin{equation}
Q^{(II)}(T\ll 1,\mu\ll d)=Q^{(I)}(T\ll 1,\mu\ll d).
\end{equation}
The desired expression for $T_{min}(d)$ can be derived by comparing the above result to $Q_{th}$. In the end one can  obtain
\begin{equation}
T_{min}^{(I)}(d)=T_{min}^{(II)}(d)\approx \frac{d\left(1-2Q_{th}\right)}{p\left(Q_{th}-\frac{e}{2}\right)},
\label{eq:Tmin}
\end{equation}
which for $p=1$ and $e=0$ becomes identical to the formula for the minimal secure transmittance of the channel connecting Alice and Bob for the case of DV QKD performed with the ideal single-photon source, derived previously in \cite{Lasota13}. 

Finding an analogous expression for $T_{min}$ for the modified version of the thermal bath DV QKD model, described in Sec.\,\ref{Sec:SPDC}, is more problematic. Due to the non-zero probability of multiphoton signal emission, in order to do this in the general case one has to derive the solution to the equation
\begin{equation}
0=y(T)\left[1-H\left(\frac{Q(T)}{y(T)}\right)\right]-H\left(y(T)\right),
\label{eq:minTequation}
\end{equation}
emerging from the formula (\ref{eq:DeltaIreal}). However, solving this equation analytically is usually impossible even with the assumptions that $T,\nu\ll 1$ and $\mu\ll d$, which simplify the expressions (\ref{eq:y}) and (\ref{eq:qber3}) for the fraction of genuine single photon pulses among all the pulses registered by Bob and the QBER to
\begin{equation}
y(T\ll1,\nu\ll 1,\mu\ll d)\approx\frac{\nu T+2d\nu-\frac{\nu^2}{2}}{\nu T+2d\nu}
\label{eq:yapprox}
\end{equation}
and
\begin{equation}
Q^{(III)}(T\ll1,\nu\ll 1,\mu\ll d)\approx\frac{\frac{e}{2}T+d}{T+2d}
\label{eq:QBER3approx}
\end{equation}
respectively. Nevertheless, derivation of an analytical expression for the minimal secure transmittance of the channel connecting Alice and Bob turns out to be possible in the two extreme situations: when $\nu\ll d$ and when $d\ll \nu$.

In the former one of these two cases, from the assumption $\nu\ll d$ follows that $y\approx 1$ and one can derive the desired expression for $T_{min}$ just by comparing the formula (\ref{eq:QBER3approx}) to $Q_{th}$. It reads
\begin{equation}
T_{min}^{(III)}(\nu\ll d)\approx \frac{d\left(1-2Q_{th}\right)}{\left(Q_{th}-\frac{e}{2}\right)}.
\label{eq:Tmin3}
\end{equation}
On the other hand, when $d\ll \nu$ one can find the minimal secure transmittance of the channel by solving the equation
\begin{equation}
y_{th}=\frac{\nu T-\frac{\nu^2}{2}}{\nu T}.
\label{eq:yth}
\end{equation}
Since QKD security requires $y<1$, when $d\ll \nu$ the inequality $d\ll T$ has to be fulfilled simultaneously. If so, $Q^{(III)}\approx e/2$ and $y_{th}$  for the formula (\ref{eq:yth}) can be found numerically from the expression
\begin{equation}
0=y_{th}\left[1-H\left(\frac{e}{2y_{th}}\right)\right]-H\left(\frac{e}{2}\right).
\label{eq:securitycondition}
\end{equation}
In this situation the final result takes the following form:
\begin{equation}
T_{min}^{(III)}(d\ll\nu)\approx \frac{\nu}{2\left(1-y_{th}\right)}.
\label{eq:Tmin4}
\end{equation}

\subsection{Minimal transmittance of the channel required for non-Gaussianity}
\label{Sec:NGminimalT}

As can be seen in Fig.\,\ref{fig:NCNGSPDCgeneral1}, in the DV QKD model with the channel noise in the form of thermal bath and SPDC-based signal source non-Gaussianity of light reaching Bob's measurement system cannot be provided for every $T$ even when $\mu=0$. In order to find the minimal transmittance of the channel connecting Alice and Bob which would be required to do this, one should begin by assuming that $\mu\rightarrow0$, $\nu\ll 1$ and $T\ll 1$ and simplifying the expressions (\ref{eq:omega13}) and (\ref{eq:omega2plus3}) in this particular situation. They take the following forms: 
\begin{equation}
\omega_{1}^{(III)}(\mu\rightarrow 0, \nu\ll1,T\ll 1)\approx T
\label{eq:omega1approx}
\end{equation}
 and
\begin{equation}
\omega_{2+}^{(III)}(\mu\rightarrow 0, \nu\ll1,T\ll 1)\approx \nu T^2/2.
\label{eq:omega2plusapprox}
\end{equation} 
Inserting (\ref{eq:omega1approx}) and (\ref{eq:omega2plusapprox}) to the  inequality (\ref{eq:nonGauss}) one can find the following expression for $T_{min}$ needed for non-Gaussianity of light arriving at Bob's side:
\begin{equation}
T_{min}^{NG,III}(\nu\ll 1)\approx \frac{\nu}{2}.
\label{eq:TminNG}
\end{equation}
For the case of $e=0$ the quantity $T_{min}^{(III)}(d\ll\nu)$ given by the formula (\ref{eq:Tmin4}) equals $T_{min}^{NG,III}(\nu\ll 1)$. This fact confirms our conclusion stated in Sec.\,\ref{Sec:bath}, that in this situation non-Gaussianity criterion remains the sufficient indication of further QKD security for every $T\ll1$ if only $d\ll T$.

\section{Summary}
\label{Sec:Summary}

We analyzed the security of DV QKD protocols over noisy channels and compared the resulting security criteria with the criteria for nonclassicality and non-Gaussianity of light reaching Bob's detectors. To do so we introduced the models for three possible physical realizations of the noisy channels, assuming that an eavesdropper fully controls the channel in each of these cases and is able to apply general collective attacks defined by the amount of the introduced errors. Our motivation has been to derive a sufficient condition for the statistics of light detected by the receiver, which would verify suitability of the link for the DV QKD. Jointly the analyzed models cover the most important types of imperfections which could possibly affect the process of key generation in realistic situations: the dark counts in Bob's detectors, the possibility for disturbing the state of signal photon during its propagation from Alice to Bob, the imperfection of realistic single-photon sources and two different kinds of channel noise. 

We found out that the nonclassicality criterion is a necessary indication of further QKD security in every realistic situation that we considered. Furthermore, when the probability for Bob to register a signal photon is small but still much higher than the probability of registering a dark count, non-Gaussianity of light arriving at his measurement system becomes a sufficient indication of further security of DV QKD protocols in the typical noisy quantum channels. In the case of Bob using modern low-noise single photon detectors the requirement of low dark count rate can be safely satisfied for a wide range of distances between the trusted parties. Both the maximal and the minimal value of the transmittance of the quantum channel connecting Alice and Bob, between which non-Gaussianity of light entering Bob's station automatically means that a given QKD scheme is secure, depend on the probability for Bob to register an error during his polarization measurement of a signal photon.

\begin{figure}
\centering
\includegraphics[width=1.0\linewidth]{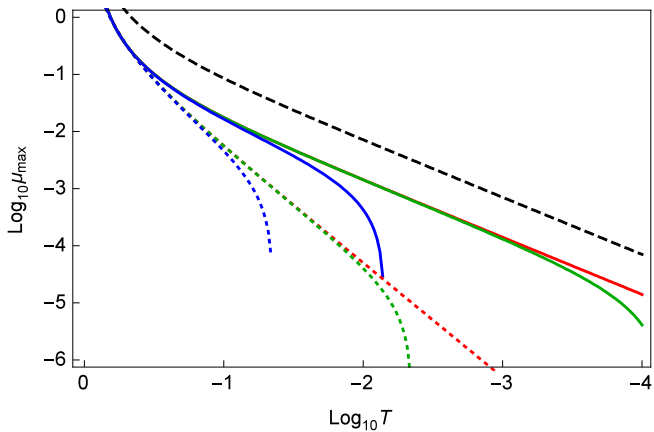}
\caption{(color online) Comparison between the lower bounds for the security of BB84 protocol (solid lines) realized using the scheme presented in Fig.\ref{fig:thermalbath} and the respective criteria for non-Gaussianity of light reaching Bob's detection system (dotted lines), calculated numerically for $p=1$ and $e=0$ for the case when his autocorrelation function measurement is performed with the same noisy and untrusty single-photon detectors that are used for him for the key generation process with the probability of registering a dark count $d=0$ (red, uppermost lines), $d=10^{-5}$ (green, middlemost lines) and $d=10^{-3}$ (blue, lowermost lines). The corresponding criterion for the nonclassicality of light arriving at Bob's side (independent of the value of $d$) is drawn with black, dashed line.}
\label{fig:realNCNGdet}
\end{figure}

For the two QKD schemes with the perfect source of signal states, analyzed in Sec.\,\ref{Sec:bath} and Sec.\,\ref{Sec:before}, the whole region of parameters $T$ and $\mu$ which is secure can be approximately bounded by two simple analytical expressions: the asymptotic non-Gaussianity criterion (given by the formula (\ref{eq:mumaxNG}) or (\ref{eq:mumaxNG2}) -- depending on the specific model) and the minimal secure value of the quantum channel transmittance (given by the formula (\ref{eq:Tmin})). It is clearly visible in Fig.\,\ref{fig:NCNGSPDCgeneral1}. The above conclusion is valid both for the case of channel noise in the form of thermal bath and for the case when the noise couples to the signal before it enters quantum channel. The similarity of the results of our analyses of both of these two situations strongly suggest that it can be valid for any type of channel noise. On the other hand, in the case of the scheme with SPDC-based signal source, analyzed in Sec.\,\ref{Sec:SPDC}, a similar asymptotic non-Gaussianity criterion, given by the expression (\ref{eq:mumaxNG3}), can be used to bound the secure region of the parameters $T$ and $\mu$, but in the general case the formula for the minimal secure $T$ cannot be derived analytically. This quantity can be found numerically by solving the equation (\ref{eq:minTequation}) with the functions $y(T)$ and $Q(T)$ given by (\ref{eq:yapprox}) and (\ref{eq:QBER3approx}) respectively. Only for the extreme situations when $\nu\ll d$ or $d\ll\nu$ simple analytical formulas for the minimal secure transmittance of the channel connecting Alice and Bob, given by the expressions (\ref{eq:Tmin3}) and (\ref{eq:Tmin4}), can be found. 
 
The results of our work show that in many situations checking non-Gaussianity of light arriving at Bob's detection system is feasible for prior assessment of the security of prepare-and-measure DV QKD systems for all elements of the tranmission link. In regime when polarization noise and dark-count rate of the detectors are small, quantum non-Gaussianity becomes a relevant {\em partial} condition on the output quantum statistics of single-mode light, which, differently to nonclassicality, sufficiently predicts the security of a DV QKD protocol once it is implemented over the given noisy quantum channel. This conclusion remains valid even for single-photon sources with non-unity probability of emission. However, the full verification of security requires further estimation of other parameters of the protocol, relevant for the security (in particular the level of QBER and the probability for Bob's measurement system to register a click when Alice's source emits a signal). It is possible that other directly measurable characteristics could be defined, which together with non-Gaussianity would form a full set of completely sufficient security indicators. However, a problem of finding such a set remains open for further research.

\medskip
\noindent {\bf Acknowledgments. --} The research leading to these results has received funding from the EU FP7 under Grant Agreement No. 308803 (project BRISQ2), co-financed by M\v SMT \v CR (7E13032). M.L. and V.C.U. acknowledge the project 13-27533J of the Czech Science Foundation. R.F. acknowledges project GB14-36681G of Czech Science Foundation. M.L. acknowledges support by the Development Project of Faculty of
Science, Palacky University.

\appendix

\section{}
\label{Sec:RealDetection}

The whole analysis presented in the main body of this article was performed with the assumption that Bob uses perfect single-photon detectors for his autocorrelation function measurements, needed for the verification of nonclassicality and non-Gaussianity of light arriving at his side. This scenario was adopted by us primarily because we wanted to check how QKD security criterion can be related to the genuine quantum features of the signal pulses reaching Bob, undisturbed by the potential imperfection of the realistic process of single-photon detection. However, in the most unfavorable situation from Bob's point of view his detectors can be noisy and untrusted, meaning that all of the errors generated by the dark counts during the process of autocorrelation function measurement have to be treated as imperfections of the detected signal. 

To check how the relationship between QKD security criterion and the criteria for nonclassicality and non-Gaussianity of light arriving at Bob's side would change if we chose this pessimistic assumption for our analysis, we repeated the calculations performed in Sec.\,\ref{Sec:bath} for the QKD model with thermal bath in the case when Bob uses exactly the same detectors for his nonclassicality and non-Gaussianity measurements as for the generation of cryptographic key (\emph{i.e.} with the same dark count probability per gate). The results of this analysis, presented in Fig.\,\ref{fig:realNCNGdet}, indicate that in this particular situation a given QKD scheme is always secure for $T\ll 1$ if only the raw results of the autocorrelation function measurement (\emph{i.e.} the results read out directly from the detectors without any correction for the expected contribution from the dark counts) indicate non-Gaussianity of the signal states reaching Bob, regardless of the value of $d$.

\section{}
\label{Sec:ThermalPoisson}

\begin{figure}
\centering
\includegraphics[width=1.0\linewidth]{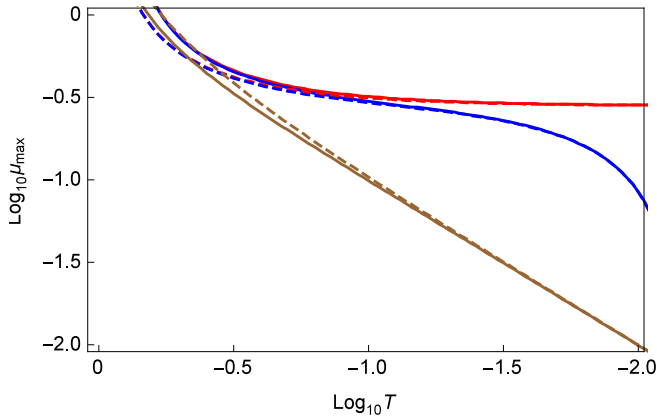}
\caption{(color online) Comparison between the lower bounds for the security of BB84 protocol realized with the use of the scheme presented in Fig.\,\ref{fig:noisebefore} and the corresponding non-Gaussianity criteria (brown lines, lowermost when $T\ll1$) for the light arriving at Bob's side, found numerically for the cases when the source of noise has thermal (solid lines) or Poisson (dashed lines) statistics of the number of photons emitted per pulse. All the calculations were performed for $p=1$ and $e=0$. The QKD security criteria were found for two different values of the probability of registering a dark count in Bob's detectors per gate: $d=0$ (red lines, uppermost when $T\ll1$) and $d=10^{-3}$ (blue lines, middlemost when $T\ll1$).}
\label{fig:NCNGnoisebeforechannel4}
\end{figure}

To check how changes of the statistics of photons emitted by a given source of noise may influence the relationship between QKD security and non-Gaussianity criteria for the DV QKD model with the noise coupling to the signal before the entrance to the quantum channel, illustrated in Fig.\,\ref{fig:noisebefore}, we performed numerical calculations for this setup configuration both for the case of Poisson and thermal types of the aforementioned statistics. The comparison of the obtained results can be seen in Fig.\,\ref{fig:NCNGnoisebeforechannel4}.

The main conclusion that can be derived from this picture is that while for high values of $T$ non-Gaussianity criterion is more restrictive in the case of thermal statistics, QKD security criterion turns out to be more relaxed in this situation. As a result of this, when the source of noise has Poisson statistics non-Gaussianity criterion is not sufficient for the QKD security for significantly broader range of high transmittance of the channel connecting Alice and Bob than in the thermal case. In fact, it may even be not sufficient when $e=0$ if $T\approx 1$, as can be seen in Fig.\,\ref{fig:NCNGnoisebeforechannel4}. However, the difference between the results of our calculations obtained for the cases of thermal and Poisson statistics of the source of noise quickly vanishes with the lowering $T$ and becomes unnoticeable when $T<10^{-1}$. It means that in the situations when the distance between Alice and Bob is long, the type of the statistics of the source of noise is not important from the perspective of security of the DV QKD model pictured in Fig.\,\ref{fig:noisebefore}.

\bibliography{refsNCNG}

\end{document}